\documentstyle[12pt,epsfig]{article}
\begin{document}
\begin{center}
{\Large{\bf{The Flavour-Changing
Neutral Currents Decay $t \rightarrow Hq$ at ATLAS Experiment}}}
\end{center}
\   \par
\   \par
\begin{center}
{\large{\bf{Leila Chikovani}}}
\end{center}
\begin{center}
Institute of Physics of the Georgian Academy of Sciencies\par
\end{center}
\   \par
\   \par
\begin{center}
{\large{\bf{Tamar Djobava}}}
\end{center}
\begin{center}
High Energy Physics Institute, Tbilisi State University\par
\end{center}
\  \par
\  \par
\begin{center}
{\large{\bf{Tamaz Grigalashvili}}}
\end{center}
\begin{center}
Institute of Physics of the Georgian Academy of Sciencies\par
\end{center}
\   \par
\   \par
\begin{center}
{\bf{ABSTRACT}}
\end{center}
\  \par
\  \par
The sensitivity of the ATLAS experiment to the
top-quark rare decay via flavour-changing neutral currents 
$t \rightarrow Hq$ ($q$ represents $c$ and $u$ quarks) has been studied
at $\sqrt{s}$=14 TeV in the decay mode of
$t\bar{t} \rightarrow HqWb \rightarrow WW^{*}q,Wb \rightarrow l \nu l \nu j,
 l^{\pm} \nu b$,
(l=e, $\mu$). 

The Standard Model backgrounds $t\bar{t}$, $t\bar{t}H$, $WZ$ and $WH$ 
 have been analysed. 
The signal and backgrounds were generated via PYTHIA 5.7 and simulated  
and analysed using ATLFAST 2.51. A branching ratio for 
$t \rightarrow Hq \rightarrow WW^{*}q$ 
 as low as 2.0x10$^{-3}$   for $m_{H}=150~ GeV$ and 2.1x10$^{-3}$ for
 $m_{H}=160~ GeV$
could be discovered at the 5$\sigma$ level with an integrated 
luminosity of 100 fb$^{-1}$.

\newpage

\section{Introduction}

 We study 
the sensitivity of the ATLAS experiment to the branching ratio
of the top-quark rare decay via flavour changing neutral currents 
(FCNC) $t \rightarrow Hq$ ($q=u,c$), where $H$ is the Standard Model Higgs 
($m_{t}$= 175 GeV, $M_{Z} \leq M_{H} \leq 2M_{W}$). 
  The SM prediction for the branching ratio is of order
Br$(t \rightarrow H_{SM}q) \approx 0.9 \cdot 10^{-13}(4 \times 10^{-15})$
for $m_{H}$=100(160) GeV [1].  
Thus, an observation of this decay mode at the LHC,
would provide a clear signal of new physics beyond the SM, such as new 
dynamical interactions of top quark,
multi-Higgs doublets, exotic fermions or other possibilities [2,3,4,5].
The recent theoretical estimations of branching is 
Br$(t \rightarrow Hq) \sim 10^{-5}\div 10^{-4}$ [5].
We consider the following process
$t\bar{t} \rightarrow HqWb \rightarrow WW^{*}q, Wb \rightarrow l\nu l\nu j,
l^{\pm} \nu b$, (l=e, $\mu$) for Higgs masses $m_{H}$=150 GeV and
$m_{H}$=160 GeV in the leptonic decay topology of W bosons since in this mass range
$H \rightarrow WW^{*}$ is the dominant decay mode.



\section{Monte Carlo Event Generation}

It is known, that the dominant mechanism
of top quark production at the LHC is $t\bar{t}$ pair production via
$gg, q\bar{q} \rightarrow t\bar{t}$.

For the implementation of the $t \rightarrow Hq$ process into PYTHIA 5.7
all individual decay channels of the top quark were switched off,
except two decay modes $t \rightarrow Wb$ and $t \rightarrow Ws$. 
The last one was replaced by the decay of the $t \rightarrow Hq$, 
by replacing of  $W$ by $H$ and $s$ by $c(u)$.

The $t\bar{t}$  events have been generated by
PYTHIA 5.7
at $\sqrt{s}=14$ TeV, $m_{top}=175$ GeV for two masses of Higgs
 $m_{H}=150$ GeV and  $m_{H}=160$ GeV with proton structure
function CTEQ2L. Initial and final state QED and QCD (ISR, FSR) radiations,
multiple interactions, fragmentations and decays of unstabled particles
were enabled. The total cross-section for
$t \bar t$ production was assumed to be $\sigma_{t\bar{t}}$=833 pb 
(NLO prediction) [6].

The following SM backgrounds have been considered for above mentioned process:\\  
$\bullet$ ~~  
 $t\bar{t} \rightarrow WbWb \rightarrow l^{+} \nu b l^{-}\bar{\nu} b$ \\
$\bullet$ ~~ 
$t\bar{t}H \rightarrow W^{+}bW^{-}\bar{b}, WW^{*} 
 \rightarrow  l^{+} \nu b, l^{-}\bar{\nu}\bar{b}, l^{+} \nu
l^{-}\bar{\nu}; l^{\pm} \nu b,jjb, l^{+} \nu l^{-}\bar{\nu}; 
l^{+} \nu b,l^{-}\bar{\nu} b,jjl\nu$\\   
$\bullet$ ~~      
$WZ \rightarrow l^{\pm} \nu  l^{+}l^{-} + X$\\
$\bullet$ ~~         
$WH \rightarrow l^{\pm} \nu WW^{*} \rightarrow l^{\pm} \nu, l^{+} \nu  
l^{-}\bar{\nu}+ X$ 

In case of $t\bar{t}$ background the third lepton originates from a semileptonic
 $b$ -decay.

All backgrounds were generated by PYTHIA 5.7.
 
The performance of the ATLAS detector was simulated using the
fast simulation package ATLFAST 2.51 [7], which uses parametrizations
of the detector resolution functions. High luminosity (KEYLUM=2) option has
been invoked. 

The b tagging perfomance was simulated assuming the nominal efficiencies of 
$\epsilon_{b}=50\%$, $\epsilon_{c} =10\%$, $\epsilon_{j} =1\%$.
The lepton identification efficiency is 
$\epsilon^{l}$ =0.9.
The branching ratios of $H \rightarrow WW^{*}$ decays have been estimated by 
the Fortran code HDECAY [8]. Br($H \rightarrow WW^{*}$)=0.6852 (0.9160) for
$m_H$=150 GeV (160 GeV).

\section{Event Analysis}

The experimental signature of $t\bar{t} \rightarrow HqWb \rightarrow WW^{*}j,
Wb \rightarrow l \nu l \nu j, l \nu b$ includes three isolated leptons, missing transverse momentum due to neutrinos, one b-jet and an additional light jet.

The leptonic decay mode of $W$ bosons have been considered for the final topology of
backgrounds.

Numbers of generated and expected events of signal and backgrounds for two masses of
 Higgs $m_{H}=150$ GeV and  $m_{H}=160$ GeV  are presented in Tables 1$\div$2.
In order to discriminate the signal from the background processes,
 the following 
preselection cuts have been applied:\\
$\bullet$~~ At least three isolated charged leptons
(electrons with $P_{T} > 5 $ GeV, $|\eta| < 2.5$ and muons with
$P_{T} > 6 $ GeV, $|\eta| < 2.4$) are required. Lepton isolation criteria
in terms of the distance from other clusters $\Delta R > $0.4 and of maximum
transverse energy deposition in cells $E_{T} <$ 10 GeV in a
cone $\Delta R$=0.2 around the leptons were applied.
\\
$\bullet$~~ Events without at least two jets with $P_{T}^{jet} > 15 $ GeV and $|\eta| < 5$,
are rejected.

The preselection cuts reduce  backgrounds  approximately from $2\%$
up to $14\%$, meanwhile 
$\sim 50\%$ of signal events are survived. 

Then a kinematical cut was applied, 
 requiring the presence of three isolated charged
leptons with high $P_{T}^{l} > 30 $ GeV. 
 This cut certainly includes the presence of one opposite charged lepton
in event.
 Such  requirement affects (reduces)
significantly
$t\bar{t}$ dangerous background, while keeping still a significant fraction of the
signal.

The next requirement of the missing transverse momentum in event with
$P_{T}^{miss} > 45$ GeV is a powerful cut for reduction of
$WZ$ background (see Fig. 1), though other backgrounds are less sensitive 
to this cut. One can see from Fig. 1, that $t\bar{t}H$ background is extended 
to larger values of $P_{T}^{miss}$ than other backgrounds due to four 
neutrinos in  $t\bar{t}H$.

We demand the presence of at least two  jets with high
$P_{T}^{jet} > 30$ GeV in the pseudorapidity  region 
$|\eta^{jet}| < 2.5$,  which satisfy the following isolation conditions: 
${\Delta}R_{jj} > 0.4$ (jet-jet isolation) and ${\Delta}R_{lj} > 0.4$ 
(lepton-jet isolation). Among the isolated high $P_{T}$ jets is required 
the presence at least of one tagged b-jet. This cut significantly suppresses 
$WZ$ background and vanishes $WH$.

For the dilepton (coming from $H$ decay) mass $m_{ll}$ reconstruction 
it have been required opposite charged
lepton pairs with $\Delta \phi <$ 1.0 ( the opening angle between the two
leptons in the transverse plane, measured in rad.), $|\Delta \eta| <$ 1.5 ( the absolute
values of the pseudorapidity difference between the two leptons) and an invariant
dilepton mass smaller than 80 GeV [9]. The small angular separation between leptons
in signal events results from the opposite spin orientation of the $W$ pair
originating from the decay of the scalar Higgs boson [10].
The discrimination between the signal and the most important backgrounds
 ($t\bar{t}$, $t\bar{t}H$) is shown for the pseudorapidity difference
 $|\Delta \eta|$ and the azimuthal difference $\Delta \phi$ between the two leptons,
 respectively, in Figures 2 and 3. 
One can see, that these cuts 
sufficiently reduce $t\bar{t}$ background and $WZ$ backhground
is vanished (see Tables 1$\div$2).
 
 In Fig. 4 are presented the dilepton invariant mass $m_{ll}$
 distributions for signal and $t\bar{t}H$, $t\bar{t}$ background events for 
a Higgs mass of 160 GeV. The solid histogram is for best combinations 
of dilepton pairs for signal and dashed histogram is for Higgs decay 
product leptons at parton level. The best combinations of $ll$ invariant 
mass is defined as the closest values to the mean of the $ll$ dilepton 
mass distribution of signal at parton level. One can see a good agreement 
between parton level and best combinations of dilepton pairs. The $t\bar{t}$ 
and $t\bar{t}H$ events are reduced by $\approx 50\%$.

Then the kinematical cuts have been applied on the two invariant masses of:\\
a) $m_{llj} < 110$ GeV with light jets $P_{T} > 30$ GeV and \\
b) $m_{lb} < 140$  GeV with $P_{T}^{bjet} > 40$ GeV.

In Fig. 5 are presented  the invariant $llj$ mass distributions for a
Higgs mass of 160 GeV for the best combinations of signal (solid 
histogram), for  $t\bar{t}H$ background (dotted histogram), $t\bar{t}$ background
(dashed-dotted histogram) and invariant mass of $llq$ for signal at 
parton level (dashed histogram). One can see  an agreement between parton
level and the best combinations of $llj$.
 
The sequence of $m_{llj} < 110$ GeV  and  $m_{lb} < 140$  GeV cuts significantly
suppresses the $t\bar{t}H$ and $t\bar{t}$ backgrounds and other remained 
backgrounds are vanished.

In Fig. 6 are presented  the invariant $lb$ mass distributions for a
Higgs mass of 160 GeV for the best combinations of signal (solid 
histogram), for  $t\bar{t}H$ background (dotted histogram), $t\bar{t}$ background
(dashed-dotted histogram) and invariant mass of $lb$ for signal at 
parton level (dashed histogram). One can see  an agreement between parton
level and the best combinations of $lb$ for the signal and remained 
backgrounds.


The acceptances and the expected rates for signal
and background events after applying all above mentioned kinematical cuts
are summarised in Tables 1$\div$2 for $m_{H}$=150 and 160 GeV.
The sensitivities to Br($t \rightarrow Hq$) have been calculated similarly
 as in [11].
Table 3 summarizes the branching ratios for
the $t \rightarrow Hq \rightarrow WW^{*}q$ decay at the different levels
 of applied cuts. 
One can see from Table 3, that
Br($t \rightarrow Hq$) as low as 2.0x10$^{-3}$ could be discovered at the 
5$\sigma$ level  for $m_{H}$=150 GeV
and as low as 2.1x10$^{-3}$ for $m_{H}$=160 GeV 
with an integrated luminosity of 100 fb$^{-1}$.

\section{Conclusions}

We have studied the ATLAS sensitivity to the FCNC top quark rare decay
$t \rightarrow Hq$ ($q=u,c$) with $H \rightarrow WW^{*}$ at
$\sqrt{s}=14$ TeV
for an integrated luminosity of $100~fb^{-1}$.  
The results demonstrate that, a branching ratio 
 as low as 2.0x10$^{-3}$  could be  discovered at the
 5$\sigma$ level  for $m_{H}$=150 GeV and  2.1x10$^{-3}$  for 
$m_{H}$=160 GeV, respectively.

\section{Acknowledgements}

We are very thankful  to M.Cobal, J.Parsons, D.Pallin and  
E.Richter-Was for very interesting and important 
discussions. 

The work was supported in part by Georgian Academy of Sciences 
grant 2.16.04.

\newpage
\begin{table}
\newcommand{\HR}
\HR\rotatebox{90}{
\begin {tabular}{|l||c|c||c|c||c|c||c|c||c|c|}\hline
\multicolumn{1}{|c||}{ Description}&\multicolumn{2}{|c||}
{$t \rightarrow Hq$}&\multicolumn{8}{|c|}{Background Processes}\\ \cline{4-11}
\multicolumn{1}{|c||}{of}&\multicolumn{2}{|c||}
{Signal}&\multicolumn{2}{|c||}{$t\bar{t}H$}&
\multicolumn{2}{|c|}{$WH$}&\multicolumn{2}{|c|}{$t \bar {t}$}&
\multicolumn{2}{|c|}{$WZ$}\\ \cline{2-11}
\multicolumn{1}{|c||}{Cuts} & Nevt & Eff (\%) & Nevt & Eff (\%) 
& Nevt & Eff (\%) & Nevt & Eff (\%)& Nevt & Eff (\%)\\ \hline \hline
Nevt gen. &10.4K& &110K& &300K & &4M& &200K&\\ \hline
Expect. events  && &12950& &430& &3.9M&&39040&\\  \hline
Preselection & 5012 & 48.00 & 395 & 3.05 & 56 & 13.07&93611&2.41&
3552&9.10\\ \hline
$P_{T}^{mis}>45~ GeV$ &  &  & 
 & & & &&&&\\ 
$P^l_{T}>30~ GeV$ & & & & & & & & & &\\
$P^{jet}_{T}>30~ GeV$  &249  & 2.39 &48  &0.38 & 0
 &$5.13\times10^{-2}$ &152& $3.91\times10^{-3}$ &5 &$1.28
\times10^{-2}$  \\
$N_{b-jet}>0$ & & & & & & & & & & \\ \hline
$m_{ll}<80~GeV$  &&& & & &
&&&&\\ 
 $\Delta\eta<1.5$ &186 &1.79 &22 &0.17 & 0&$4.03\times10^{-2}$
&38&$9.78\times10^{-4}$&0&$1.00\times10^{-3}$\\ 
 $\Delta\phi<1.0$ & & & & & & & & & & \\
\hline \hline
$t\rightarrow Hq$ & & &  & & 
& &&&& \\ 
$m_{llj}<110~ GeV$ & 87& 0.84 & 6 & $5.09\times10^{-2}$ & 0
& $7.30\times10^{-3}$&7& $1.80\times10^{-4}$&0&
 $5.00\times10^{-4}$ \\
$P^{j}_{T} > 30~ GeV$ & & & & & & & & & & \\ \hline
$t\rightarrow Wb$ &  &  &  & & 
& && & &   \\ 
$m_{lb}<140~ GeV$ & 78 & 0.75 & 6 & $4.64\times10^{-2}$ & 0
& $6.70\times10^{-3}$&3& $7.72\times10^{-5}$&0& $5.00\times10^{-4}$  \\ 
 $P^{b}_{T} > 40~ GeV$ & & & & & & & & & & \\ \hline
\end{tabular}}
\newcommand{\LT}
\LT\rotatebox{90}{
\begin{tabular}{l}
Table 1: The numbers of events and efficiencies ($\%$) of kinematic cuts
applied in sequence for the signal and backgrounds \\
for $m_{H}=150~ GeV$ (are not multiplied on the lepton identification efficiency
$(\epsilon^{l})^3$).The numbers of events after cuts for \\
backgrounds are presented according to the expected events.
\end{tabular}}
\end{table}
\newpage
\begin{table}
\newcommand{\TR}
\TR\rotatebox{90}{
\begin {tabular}{|l||c|c||c|c||c|c||c|c||c|c|}\hline
\multicolumn{1}{|c||}{ Description}&\multicolumn{2}{|c||}
{$t \rightarrow Hq$}&\multicolumn{8}{|c|}{Background Processes}\\ \cline{4-11}
\multicolumn{1}{|c||}{of}&\multicolumn{2}{|c||}
{Signal}&\multicolumn{2}{|c||}{$t\bar{t}H$}&
\multicolumn{2}{|c|}{$WH$}&\multicolumn{2}{|c|}{$t \bar {t}$}&
\multicolumn{2}{|c|}{$WZ$}\\ \cline{2-11}
\multicolumn{1}{|c||}{Cuts} & Nevt & Eff (\%) & Nevt & Eff (\%) 
& Nevt & Eff (\%) & Nevt & Eff (\%)& Nevt & Eff (\%)\\ \hline \hline
Nevt gen. &8.8K& &32.8K& &15K & &4M& &200K&\\ \hline
Expect. events  && &15020& &480& &3.9M&&39040&\\  \hline
Preselection & 3943 & 45.00 & 475 & 3.16 & 66 & 13.70&93611&2.41&
3552&9.10\\ \hline
$P_{T}^{mis}>45~ GeV$ &  &  & 
 & & & &&&&\\ 
$P^l_{T}>30~ GeV$ & & & & & & & & & &\\
$P^{jet}_{T}>30~ GeV$  &218  & 2.48 &76  &0.51 & 0
 &$5.33\times10^{-2}$ &152& $3.91\times10^{-3}$ &5 &$1.28
\times10^{-2}$  \\
$N_{b-jet}>0$ & & & & & & & & & & \\ \hline
$m_{ll}<80~GeV$  &&& & & &
&&&&\\ 
 $\Delta\eta<1.5$ &170 &1.93 &35 &0.24 & 0&$4.67\times10^{-2}$
&38&$9.78\times10^{-4}$&0&$1.00\times10^{-3}$\\ 
 $\Delta\phi<1.0$ & & & & & & & & & & \\
\hline \hline
$t\rightarrow Hq$ & & &  & & 
& &&&& \\ 
$m_{llj}<110~ GeV$ & 60& 0.68 & 12 & $8.54 \times10^{-2}$ & 0
& $6.70\times10^{-3}$&7& $1.80\times10^{-4}$&0&
 $5.00\times10^{-4}$ \\
$P^{j}_{T} > 30~ GeV$ & & & & & & & & & & \\ \hline
$t\rightarrow Wb$ &  &  &  & & 
& && & &   \\ 
$m_{lb}<140~ GeV$ & 59 & 0.67 & 11 & $7.62\times10^{-2}$ & 0
& $5.10\times10^{-3}$&3& $7.72\times10^{-5}$&0& $5.00\times10^{-4}$  \\ 
 $P^{b}_{T} > 40~ GeV$ & & & & & & & & & & \\ \hline
\end{tabular}}
\newcommand{\KT}
\KT\rotatebox{90}{
\begin{tabular}{l}
Table 2: The numbers of events and efficiencies ($\%$) of kinematic cuts
applied in sequence for the signal and backgrounds \\
for $m_{H}=160~ GeV$ (are not multiplied on the lepton identification efficiency
$(\epsilon^{l})^3$).The numbers of events after cuts for \\
backgrounds are presented according to the expected events.
\end{tabular}}
\end{table}
\newpage
\begin{table}[hbpt]
\begin {tabular}{|l||c|c|} \hline
\multicolumn{1}{|c||}{Cuts}&\multicolumn{2}{|c|}{Sensitivity to
Br(t $\rightarrow$ Hq)}\\ \cline{2-3}
 & $m_{H}=150 ~GeV$ & $m_{H}=160 ~GeV$\\
 \hline \hline
$t\rightarrow Hq$ & 2.19$\times 10^{-3}$ & 2.43$\times 10^{-3}$ \\ 
$m_{llj}<110~ GeV$, $P^{j}_{T} > 30~ GeV$ & & \\ \hline
$t\rightarrow Wb$ &  2.03$\times 10^{-3}$ &  2.12$\times 10^{-3}$ \\
$m_{lb}<140~ GeV$, $P^{b}_{T} > 40~ GeV$& & \\ \hline
\end{tabular}
\      \par
\    \par
\      \par
\    \par
Table 3: Summary of the results at $L=100~fb^{-1}$.
\label{tab:summary}
\end{table}
\newpage
\begin{figure}
\begin{center}
\epsfig{file=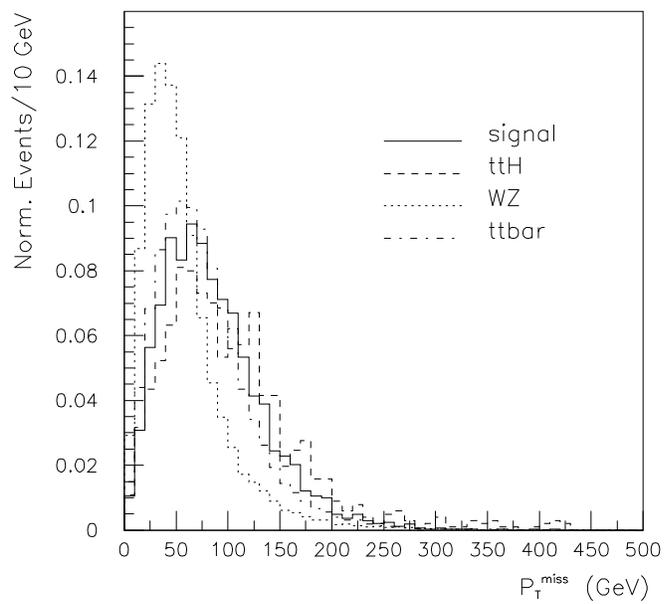,bbllx=0pt,bblly=0pt,bburx=594pt,bbury=842pt,
width=18cm,angle=0}
\end{center}
\vspace{-7.cm}
\hspace{0.cm}
\begin{minipage}{16.0cm}
\caption
{The  $P_{T}^{miss}$ distributions for signal events with
$m_{H}=160$ GeV  and $t\bar{t}H$, $t\bar{t}$ and $WZ$ backgrounds. 
All distributions are normalised to unity.} 
\end{minipage}
\end{figure}
\newpage
\begin{figure}
\begin{center}
\epsfig{file=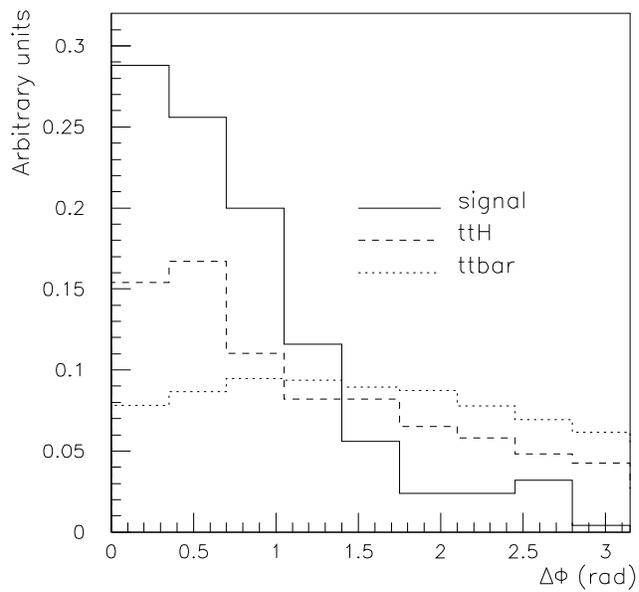,bbllx=0pt,bblly=0pt,bburx=594pt,bbury=842pt,
width=18cm,angle=0}
\end{center}
\vspace{-7.cm}
\hspace{0.cm}
\begin{minipage}{16.0cm}
\caption
{Difference in azimuth between the two leptons for signal events with
$m_{H}=160$ GeV  and $t\bar{t}$ and $t\bar{t}H$ 
 backgrounds. All distributions are normalised to unity.} 
\end{minipage}
\end{figure}
\newpage
\begin{figure}
\begin{center}
\epsfig{file=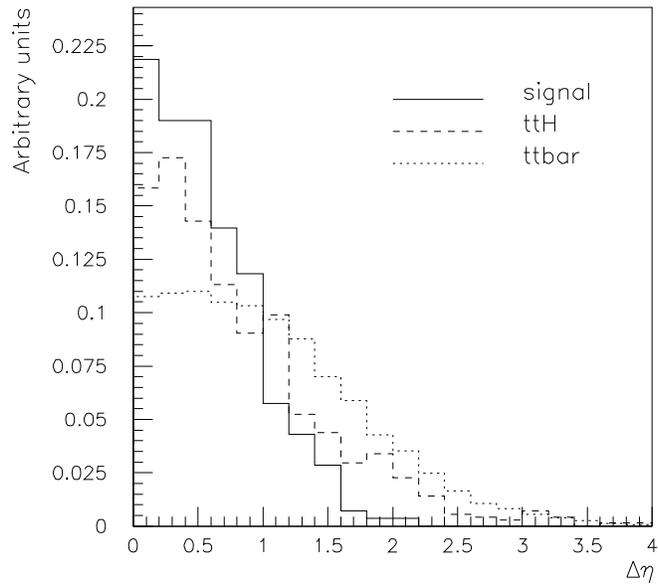,bbllx=0pt,bblly=0pt,bburx=594pt,bbury=842pt,
width=18cm,angle=0}
\end{center}
\vspace{-7.cm}
\hspace{0.cm}
\begin{minipage}{16.0cm}
\caption
{The absolute values of the
 pseudorapidity difference between the two leptons for signal events with
$m_{H}=160$ GeV  and $t\bar{t}$ and $t\bar{t}H$ 
 backgrounds. All distributions are normalised to unity.} 
\end{minipage}
\end{figure}
\newpage
\begin{figure}
\begin{center}
\epsfig{file=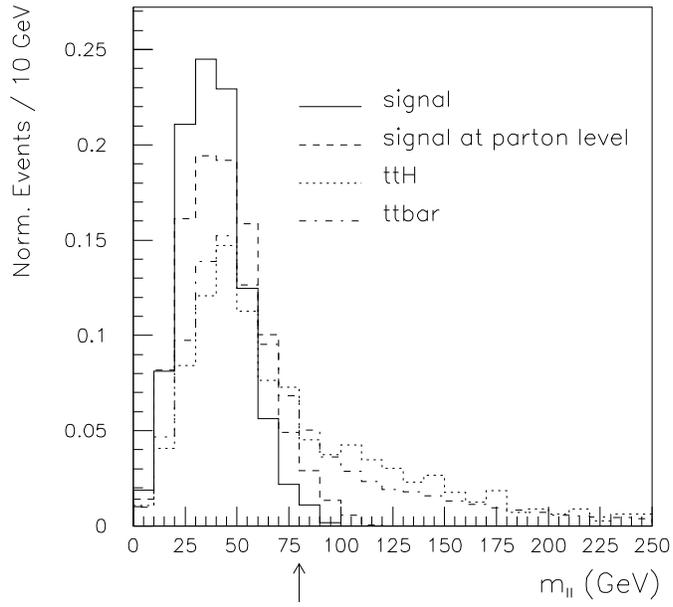,bbllx=0pt,bblly=0pt,bburx=594pt,bbury=842pt,
width=18cm,angle=0}
\end{center}
\vspace{-7.cm}
\hspace{0.cm}
\begin{minipage}{16.0cm}
\caption
{Invariant $ll$ mass distributions for the signal and background
events: the solid histogram for the best combinations of signal, the dashed 
histogram for signal at parton level, the dotted histogram for $t\bar{t}H$ 
background, the dashed-dotted histogram for $t\bar{t}$. The arrow indicates 
the value of cut on dilepton mass.}
\end{minipage}
\end{figure}
\newpage
\begin{figure}
\begin{center}
\epsfig{file=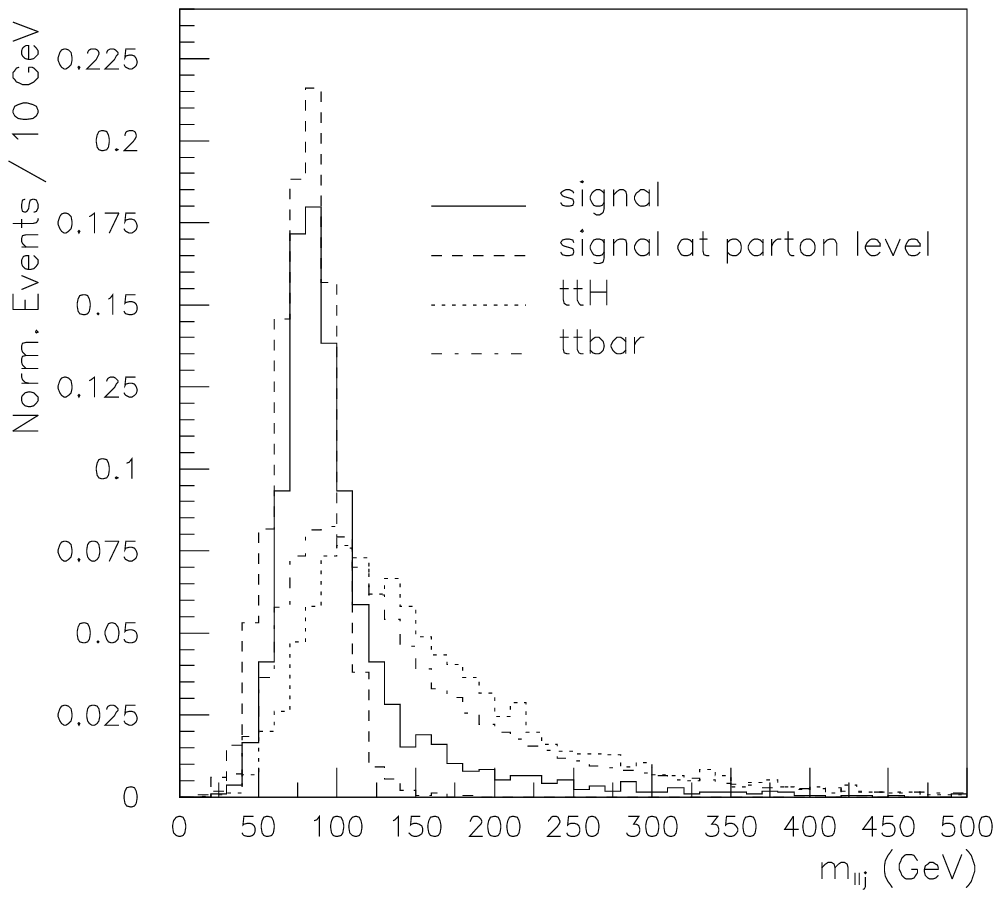,bbllx=0pt,bblly=0pt,bburx=594pt,bbury=842pt,
width=18cm,angle=0}
\end{center}
\vspace{-7.cm}
\hspace{0.cm}
\begin{minipage}{16.0cm}
\caption
{Invariant $llj$ mass distributions for the signal and background
events: the solid histogram for the best combinations of signal, the dashed 
histogram for signal at parton level, the dotted histogram for $t\bar{t}H$ 
background, the dashed-dotted histogram for $t\bar{t}$.}
\end{minipage}
\end{figure}
\begin{figure}
\begin{center}
\epsfig{file=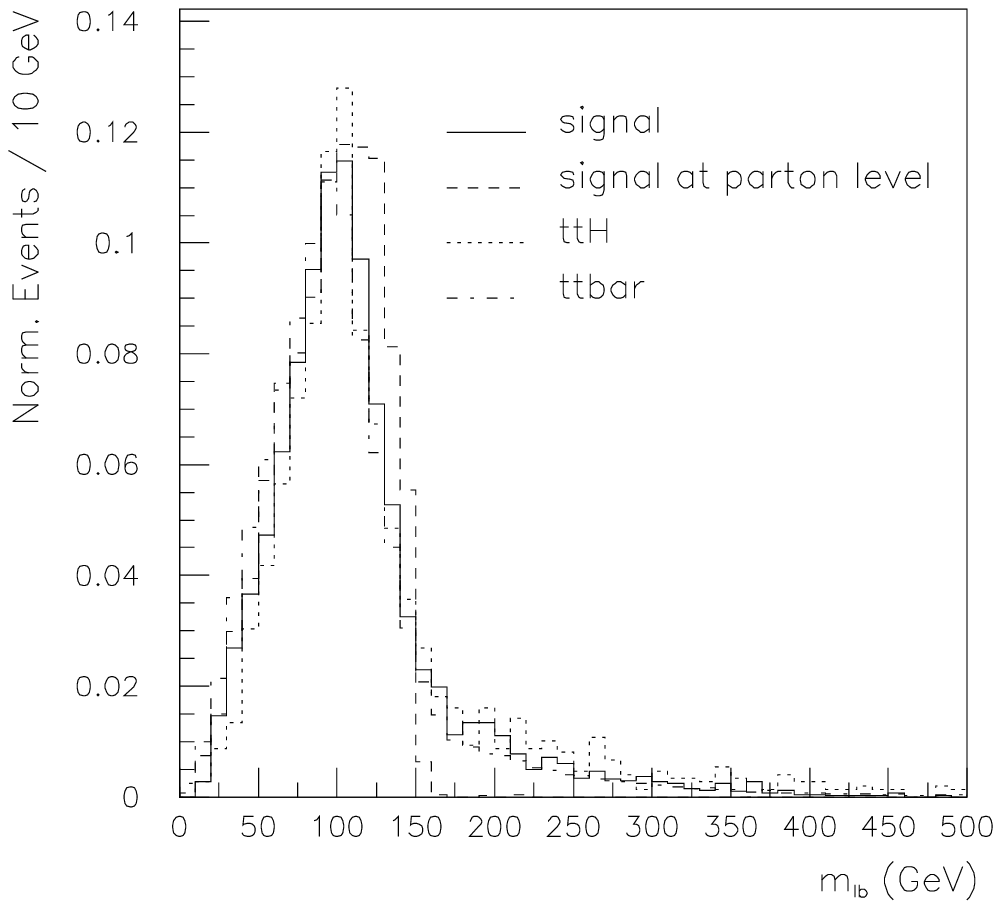,bbllx=0pt,bblly=0pt,bburx=594pt,bbury=842pt,
width=18cm,angle=0}
\end{center}
\vspace{-7.cm}
\hspace{0.cm}
\begin{minipage}{16.0cm}
\caption
{Invariant $lb$ mass distributions for the signal and background
events: the solid histogram for the best combinations of signal, the dashed 
histogram for signal at parton level, the dotted histogram for $t\bar{t}H$ 
background, the dashed-dotted histogram for $t\bar{t}$.}
\end{minipage}
\end{figure}
\end{document}